\documentclass[preprint,aps,floats,showpacs,amssymb,tightenlines]{revtex4}

\usepackage{epsfig}

\newcommand{\be}{\begin{equation}}
\newcommand{\ee}{\end{equation}}
\newcommand{\bea}{\begin{eqnarray}}
\newcommand{\eea}{\end{eqnarray}}
\newcommand{\bml}{\begin{mathletters}}
\newcommand{\eml}{\end{mathletters}}

\begin{document}

\tighten

\draft




\title{Soliton solutions of the non-linear Schr\"odinger equation
with nonlocal Coulomb and Yukawa-interactions }
\renewcommand{\thefootnote}{\fnsymbol{footnote}}

\author{Betti Hartmann\footnote{b.hartmann@iu-bremen.de}}
\affiliation{School of Engineering and Science, International University Bremen (IUB),
28725 Bremen, Germany}
\author{Wojtek J. Zakrzewski \footnote{w.j.zakrzewski@durham.ac.uk}}
\affiliation{Department of Mathematical Sciences, University of Durham,
Durham DH1 3LE, UK}
\date{\today}
\setlength{\footnotesep}{0.5\footnotesep}

\begin{abstract}
We study the non-linear Schr\"odinger equation in $(1+1)$ dimensions
in which the nonlinear term is taken in the form of a nonlocal interaction
 of the  Coulomb or Yukawa-type.
 We solve the equation numerically and find that, for all values of the nonlocal 
coupling constant, and in all cases, the equation possesses
solitonic solutions. We show that our results, for the dependence of the height
of the soliton on the coupling constant, are in good agreement
with the predictions based on an analytic
treatment in which the soliton is approximated by a gaussian.
\end{abstract}

\pacs{05.45.Yv}
\maketitle
\section{Introduction}
The Nonlinear Schr\"odinger equation (NLS) in $(1+1)$ dimensions arises
in many descriptions of physical phenomena which range from the water waves 
and wave propagation in fibre optics to biophysics \cite{bio}. Moreover,
being perhaps the simplest nonlinear PDE, it has been studied a lot
by both mathematicians and physicists. In particular the equation 
was shown to be integrable and that it has soliton-like solutions \cite{ablowitz}.
Most of this work involved the simplest version of the equation 
- in which the nonlinearity was in the form of a simple power of the 
field itself.

The equation has also been extended in many ways; people
looked at the effects of the additions of higher powers of the fields and of
their derivatives. Some of these generalisations involved also nonlocal terms.
In fact, some of them, in the optics literature, go under the name
of the Gross-Pitaevskii equation \cite{gp}. 
Due to its general nature the equation
has been used in many applications. Recently these generalisations
involving a nonlocal nonlinear term have been used in
 the description of the Bose-Einstein condensates
below the transition temperature \cite{dsps}. In this case the external potential
models the trap for the particles, while the nonlocal interaction
describes the interaction between the bosons in the condensate.
The nonlocal equation has also recently been used to describe
wave dynamics in media with a large response length, 
{\it e.g.} in nonlinear optics (see \cite{ao} and references therein).
The nonlocal equation was also studied some time ago by Ginibre and Velo \cite{ginibre}
who, in their mathematical paper, proved that some nonlocal Schr\"odinger equations
possess well defined solutions. This proof did not give any details of their
nature, {\it i.e.} it did not answer the question whether the solutions represent solitons
and what their properties are.

More recently Abe and Ogawa \cite{ao} looked at the nonlocal version of the nonlinear Schr\"odinger
equation $(1+1)$ dimensions for a class of nonlinear terms involving a power-law interaction.
They demonstrated the existence of solitonic solutions and then, using
a gaussian approximation,  studied
the dependence of the width of the solitons on the exponent of the power-law
interaction. 
However, the study did not include the cases of a Coulomb or Yukawa interactions.
In \cite{pkg} the study was extended to more than one spatial dimension
using an exponential-type nonlocal interaction. 

Given that in most applications the nonlocal term comes from the interaction of some
(spread-out) physical charges we have decided to have another look at this equation for the cases
of Coulomb - and Yukawa-like interactions. 

In this paper, we present the results of our studies
of a system in one spatial dimension with nonlocal interaction, 
{\it i.e.} when the kernel of the nonlocal term
is given by $-\kappa \vert x-y\vert$ or $\kappa {\exp(-\gamma\vert x-y\vert)\over \gamma}$.
Our paper is organised as follows: In Section II, we introduce all the notations and
 present the model. In Sections III and IV, respectively,
we discuss the cases of the Coulomb - and Yukawa interactions. In each Section we present 
the numerically obtained results for the existence of solitons and their properties
and then present some analytic
arguments based a gaussian approximation which allow us to 
explain these results. We finish with some conclusions in Section V.

\section{The model}
The action of the model in $(1+1)$ dimensions, which we have studied, is given by:
\begin{equation}
S=\int \int dt \ dx \left[ \frac{i}{2} \left(\psi^* \frac{\partial \psi}{\partial t} - \frac{\partial \psi^*}{\partial t} \psi\right)
- \left\vert \frac{\partial \psi}{\partial x}\right\vert^2 \right]\,+
\end{equation}
$$
+\frac{\kappa}{2} \int\int\int dt \ dx \ dx' \vert \psi(x,t)\vert^2 K(x,x') \vert \psi(x',t)\vert^2 
$$
and the corresponding equation reads:
\begin{equation}
\label{nnls}
i\frac{\partial \psi}{\partial t} + \frac{\partial^2 \psi}{\partial x^2} +
\kappa \left( \int K(x,x') \vert\psi(x')\vert^2 dx'\right) \psi = 0  \ . 
\end{equation}
Here $K(x,x')$ is the kernel of the non-local interaction of the system. 
For $K(x,x')=\delta(x-x')$, {\it i.e.} in the 
``local'' limit,  (\ref{nnls}) reduces to
the usual non-linear Schr\"odinger equation, which is known to possess soliton solutions,
while the case
$K(x,x')=\vert x-x'\vert^{-k}$, $0 < k < 1$ was studied in \cite{ao}.
Here, we are interested in the physically more relevant Coulomb and Yukawa-type interactions
which lie outside the range of potentials studied in \cite{ao}.

Note that the system possesses some conserved 
quantities, like the energy which is given by:
\begin{equation}
\label{hamiltonian}
H=\int\limits_{-\infty}^{\infty} dx  \left\vert \frac{\partial \psi}{\partial x}\right\vert^2
- \frac{\kappa}{2} \int\limits_{-\infty}^{\infty} dx \int\limits_{-\infty}^{\infty} dx' \vert\psi(x,t)\vert^2 K(x,x') \vert\psi(x',t)\vert^2   \ .
\end{equation}
and the intensity (or the number of particles)
\begin{equation}
N=\int \vert \psi(x,t)\vert^2 dx 
\end{equation}
which, in this work, for convenience, we have set equal to one. Often we can ``undo'' this setting of $N=1$
by an appropriate scaling of $x$, $t$ and $\psi$.

\section{Coulomb-type interaction}
First we consider a Coulomb-type interaction, {\it i.e.} the case of $K(x,x')=-\vert x-x'\vert$.

Clearly the equation (\ref{nnls}) is too hard to solve analytically. However, we expect it to have
solitonic solutions. Note that both terms in (\ref{hamiltonian}), due to 
the $-$ sign in $K$, are positive but their roles
are very different; the first term tends to spread the field while the second one tends 
to focus it. Hence we expect the interplay of both terms to lead to a stable 
(stationary) solution which minimises the Hamiltonian $H$.  

What is this solution like? Clearly, to get some intuition for its shape we can 
try a gaussian of the form:
\begin{equation}
\label{gauss}
\psi(x)=\frac{\sqrt{\alpha}}{\pi^{1/4}} \exp\left(-\frac{\alpha^2}{2}x^2\right)  \ .
\end{equation}
If this approximation is relevant then 
the height of the soliton, defined here as
the max of $\psi\psi^*$, in this Ansatz is proportional to $\alpha$, which is a free
parameter in (\ref{gauss}), while its
width is proportional to $1/\alpha$.

Inserting (\ref{gauss}) into (\ref{hamiltonian}), we find that
\begin{equation}
\label{energy_coulomb}
H= \frac{1}{2} \alpha^2 + \frac{\kappa} {\alpha \sqrt{2\pi}} \ .
\end{equation}
The value of the parameter $\alpha$ which minimises the energy then is given by
\begin{equation}
\label{coulomb_height}
\frac{\alpha}{\sqrt{\pi}}=\left(\frac{\kappa}{\sqrt{2}\pi^2}\right)^{1/3}  \ .
\end{equation}
Note that increasing strength of the nonlocal interaction $\kappa$, 
the height of the soliton (given by $\frac{\alpha}{\sqrt{\pi}}$)
increases and the width (given by $1/\alpha$) decreases. For $\kappa=0$ (the case of the
``pure'' {\it i.e.} linear Schr\"odinger equation), no localised structures appear, as expected.

Note that, as expected, (\ref{energy_coulomb}) shows that the energy
of the approximate solution consists of two terms which exhibit
a different behaviour when one varies $\alpha$.  To predict the dependence of the 
minimal energy on the coupling $\kappa$, we insert (\ref{coulomb_height}) into (\ref{energy_coulomb}) and 
find that 
\begin{equation}
\label{energy2}
H=\frac{3}{2} \frac{\kappa^{2/3}}{(2\pi)^{1/3}}   \ .
\end{equation}  
\subsection{Numerical results}

Next we have checked how good our predictions based on the gaussian approximation really are.
To do this we have performed several numerical calculations using 
a fourth-order Runge-Kutta method for simulating the time evolution.
Of course the $x$-variable has been discretised and we have typically used 500 lattice points.

\begin{figure}[!htb]
\centering
\leavevmode\epsfxsize=9.0cm
\epsfbox{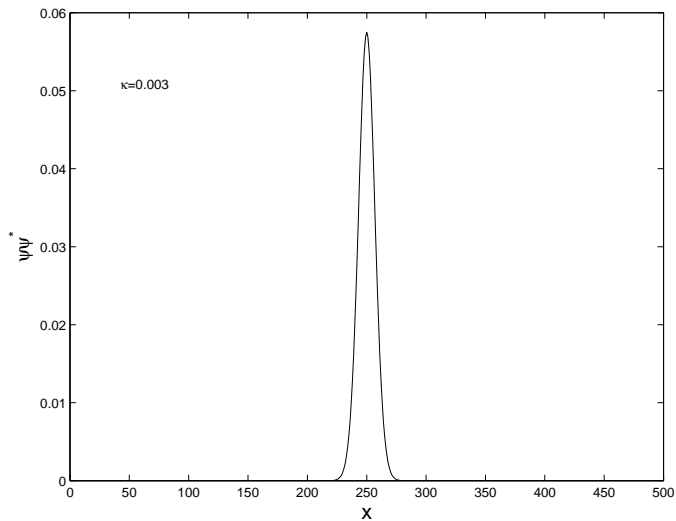}\\
\caption{\label{fig1} 
The value of $\psi\psi^*$ in dependence on $x$ is shown for a typical soliton
solution with $\kappa=0.003$. }
\end{figure}

We have used several starting field configurations.
Most of our simulations involved starting with a strongly localised structure,
{\it e.g.} $\vert\psi(x_0)\vert^2=1$, $\vert\psi(x\neq x_0)\vert^2 = 0$ with
$x_0=250$.  Then we took the energy out
of the system by adding a small dissipative term into (\ref{nnls})
having first taken most of the explicit
time dependence of $\psi(x,t)$ into account by 
writing $\psi(x,t)$ as 
\begin{equation}
\psi(x,t)\,=\,\exp(iEt)\,g(x,t),
\end{equation}
where $E$ can be calculated from (\ref{nnls}),
and really only evolving the equation for $g(x,t)$.

 In consequence, the system evolved into a
static equation for $g(x,t)=g(x)$, {\it i.e.} a stationary solution field equation for $\psi(x,t)$.

A typical soliton solution is shown in Fig.\ref{fig1} 
(it corresponds to $\kappa=0.003$). We have found that the final
stationary configurations were very well localised even for small values of $\kappa$.

\begin{figure}[!htb]
\centering
\leavevmode\epsfxsize=9.0cm
\epsfbox{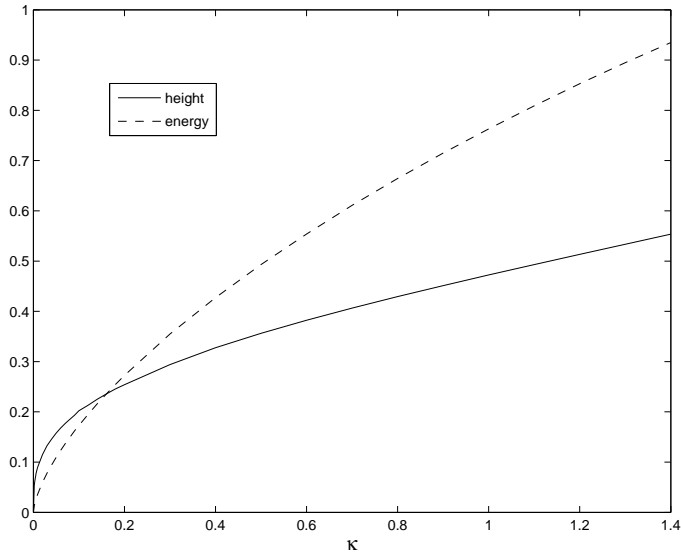}\\
\caption{\label{fig2} 
The height of the soliton $(\psi\psi^*)_{max}$  as well as its energy for a nonlocal
Coulomb interaction as a function of the nonlocal coupling $\kappa$.}
\end{figure}

Next we studied the dependence of the height and of the energy of the stationary
solution as a function of the nonlocal coupling constant $\kappa$.
 Our results are shown in Fig.\ref{fig2}.

Clearly, we have found that soliton solutions exist for all values of the nonlocal
coupling $\kappa$.

Fitting our numerical data for the height of the soliton, we see that $(\psi\psi^*)_{max} \approx 0.4732 \cdot 
\kappa^{0.3713}$ for $\kappa \in [0:1.4]$ and  $(\psi\psi^*)_{max} \approx 0.4372 \cdot 
\kappa^{0.3394}$ for $\kappa \in [0:0.1]$. For the energy, we have
$H=0.7608 \cdot \kappa^{0.6390}$ for  $\kappa \in [0:1.4]$ and
$H=0.7968 \cdot \kappa^{0.6630}$ for  $\kappa \in [0:0.1]$. Thus for small values of $\kappa$ our numerical
results agree extremely well with the predictions
based on the gaussian approximation - given by the expressions in (\ref{coulomb_height}) and (\ref{energy2}).

\section{Yukawa-type interaction}
Next we have looked at a Yukawa-type interaction, for which we have $K(x,x')=\frac{1}{\gamma}\exp(-\gamma\vert x-x'\vert)$.
Note that $\gamma=0$ corresponds to the case of the Coulomb-interaction studied above.

This time it is even harder to get any analytical results.
So, again we look at a gaussian approximation.
Thus we insert (\ref{gauss}) into (\ref{hamiltonian}) and find that
\begin{equation}
H= \frac{1}{2} \alpha^2 -  \frac{\kappa}{2\gamma\pi} {\cal I}\left(\frac{\gamma}{\alpha}\right)
\end{equation}
with
\begin{eqnarray}
{\cal I}\left(\frac{\gamma}{\alpha}\right) &=& \frac{\sqrt{\pi}}{2}\int\limits_{\infty}^{\infty} dx' \  \exp
\left[-x'^2-\frac{\gamma}{\alpha} x'+\frac{1}{4}(\frac{\gamma}{\alpha})^2 \right] \left(1+ 
\exp\left[2\frac{\gamma}{\alpha} x'\right] \nonumber \right.\,+ \\
&+& \left. {\rm erf}\left[x'-\frac{\gamma}{2\alpha}\right] - {\rm erf}\left[x'+\frac{\gamma}{2\alpha}\right]
\exp\left[2\frac{\gamma}{\alpha} x'\right]\right),
\end{eqnarray}
where erf is the errorfunction. The value of this integral has to be evaluated
numerically. We find that {\it e.g.} ${\cal I}(1)\approx 1.643$, ${\cal I}(2)\approx 1.056$,
${\cal I}(1/8)\approx 2.851$, {\it  i.e.} we see that for fixed $\gamma$, the value of the integral increases
with increasing $\alpha$. The energy thus decreases with the increasing height of the approximate soliton.

To be able to give arguments on how the height of the soliton
depends on the coupling $\gamma$, we expand the exponential for small $\gamma$ as
\begin{equation}
\frac{1}{\gamma} \exp(-\gamma \vert x-x'\vert) \approx \frac{1}{\gamma} - 
\vert x-x'\vert + \frac{\gamma}{2} \vert x-x'\vert^2 + O(\gamma^2)  \ .
\end{equation}
Neglecting $O(\gamma^2)$-terms, we obtain for the energy
\begin{equation}
\label{energy_yukawa}
H= \frac{1}{2} \alpha^2 - \frac{\kappa}{2\gamma} + \frac{\kappa}{\alpha \sqrt{2\pi}} - \frac{\gamma \kappa}{4\alpha^2}
\end{equation} 
where the second term on the rhs is a constant piece that is not important for the arguments that follow.
The value of $\alpha$ that minimises the energy fulfills the equation
\begin{equation}
\label{eq2}
\alpha^4 - \frac{\kappa}{\sqrt{2\pi}} \alpha + \frac{\gamma \kappa}{2} = 0 \ .
\end{equation}
Since the expressions for the roots of this equation are complicated, we evaluate the required value
of $\alpha$ by comparing it to the solution in the case when $\gamma=0$ {\it i.e.}
 $\alpha_0=\left(\frac{\kappa}{\sqrt{2\pi}}\right)^{1/3}$.
Thus we put $\alpha=\alpha_0+ \rho \gamma$ and inserting this into (\ref{eq2}) we find that
\begin{equation}
\label{yukawa_height}
\alpha= \alpha_0 - \frac{\sqrt{2\pi}}{6} \gamma \ \ \ {\rm for} \ \ \gamma \ {\rm small}  \ .
\end{equation}
This tells us that for fixed values of $\kappa$, the height of the soliton
 {\it decreases} with $\gamma$. Moreover, in the linear approximation
in $\gamma$, the correction to the
height doesn't depend on the strength of the nonlocal coupling $\kappa$.

\subsection{Numerical results}
\begin{figure}[!htb]
\centering
\leavevmode\epsfxsize=9.0cm
\epsfbox{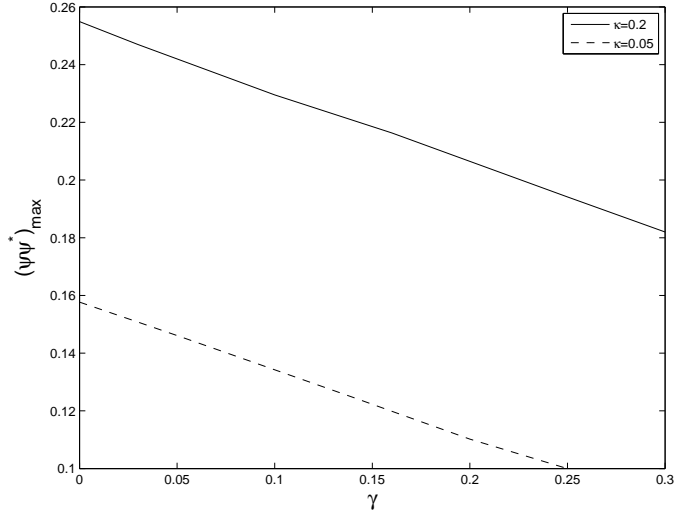}\\
\caption{\label{fig3} 
The height of the soliton $(\psi\psi^*)_{max}$ 
for a non-local Yukawa interaction as a function of $\gamma$ for $\kappa=0.2$ and $\kappa=0.05$, respectively.}
\end{figure}

Again, as in the Coulombic case,
we have studied this problem numerically using the same procedures as before.
In all studied cases we have found solitonic - like stationary solutions.
Next we have looked at the properties of these solutions. 
We have fixed $\kappa$ and have tried to determine the height and the energy of the 
stationary soliton as a function of the parameter $\gamma$. We have found that in all
cases the soliton height decreases as one increases $\gamma$.
In Fig.\ref{fig3} we present our results for $\kappa=0.2$ and $\kappa=0.05$, respectively.
As predicted by our analytic argument
(\ref{yukawa_height}), we observe that the height of the soliton decreases linearly with $\gamma$
and that the two curves are parallel to each other indicating the independence of the decrease on $\kappa$.

\begin{figure}[!htb]
\centering
\leavevmode\epsfxsize=9.0cm
\epsfbox{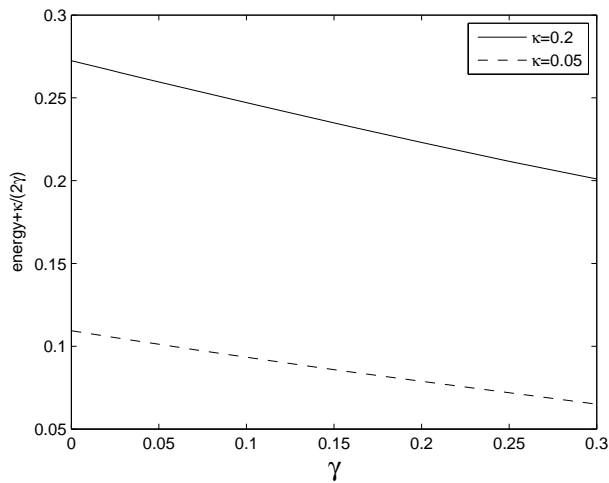}\\
\caption{\label{fig4} 
The energy of the soliton 
for a non-local Yukawa interaction as a function of $\gamma$ for $\kappa=0.2$ and $\kappa=0.05$, respectively.
Note that we have subtracted 
the term $\kappa/{2\gamma}$ from the energy (compare (\ref{energy_yukawa})).}
\end{figure}

In Fig.\ref{fig4}, we show the energy of the solutions as a function of $\gamma$ for $\kappa=0.2$ and $\kappa=0.05$.
Note that when we have added the constant piece $\kappa/(2\gamma)$ to the energy (compare also
(\ref{energy_yukawa})), in order to make the
energy finite for $\gamma=0$, the energy also decreases.
For $\gamma=0$, the Yukawa interaction reduces to the Coulomb interaction, while for $\gamma\to \infty$
we obtain the case of the ``pure'' Schr\"odinger equation. We would thus expect the
height and the rescaled energy to tend to the values found in the Coulomb case for $\gamma\to 0$ and our numerical
results confirm this. Moreover, for $\gamma\to\infty$, there is no localised structure and so
the height and the energy, as expected, tend to zero.

\section{Conclusions}
We have studied a class of nonlinear Schr\"odinger models in which the nonlinear term
involves nonlocal interactions based on  Coulomb or Yukawa like potentials.
We have shown that, like the more familiar local nonlinear Schr\"odinger model, these models
also possess solitonic-like stationary solutions for all values of the coupling constant.
All solutions in the Yukawa case are less localised than the corresponding
Coulombic solutions. As expected, when one increases the strength of the potential 
the solitons become more localised.

The solitonic fields look a little like gaussians; in fact an approximation of them by
a gaussian is very reliable. We have calculated various properties of the solitons
based on this approximation and have found them to be in a good agreement with what we have seen
in our simulations of the full equations.

In \cite{yves} a similar gaussian approximation was successfully applied to the problems involving
a local nonlinear term (but involving derivatives) in many dimensions. Would this be the case
when we look at nonlocal terms in two or even three spatial dimensions?
This problem is currently being looked at.\\
\\

{\bf Acknowledgements} WJZ thanks the ICTS visitors programme of the International
University Bremen for some financial support which made possible his visit to Bremen.
We also thank Niels Walet for providing us with references of papers on graphene - which
indirectly led to the the undertaking of this study.


\end{document}